\documentstyle[aps,epsfig,multicol]{revtex}
\renewcommand{\narrowtext}{\begin{multicols}{2} \global\columnwidth20.5pc}
\renewcommand{\widetext}{\end{multicols}\global\columnwidth42.5pc}
\begin{document}

\title{INTERFACIAL TENSION IN WATER AT SOLID
SURFACES\footnote{Published in proceedings of the \emph{Third
International Symposium on Cavitation},
vol. {\bf 1}, p87-90 (1998).}}

\author{N.A. Mortensen, A. K\"{u}hle and K.A. M\o
rch\footnote{Corresponding author. E-mail: \emph{K.A.Morch@fysik.dtu.dk}}}

\address{Department of Physics, Bldg. 307, Technical University of
Denmark, DK-2800 Lyngby, Denmark} 
\maketitle

\begin{abstract}
A model for the formation of cavitation nuclei in liquids has recently
been presented with basis in interfacial liquid tension at non-planar
solid surfaces of concave form. In the present paper investigations of
water-solid interfaces by atomic force microscopy are reported to
illuminate experimentally effects of interfacial liquid tension. The
results support that such tension occurs and that voids develop at solid-liquid interfaces.
\end{abstract}

\narrowtext

\section*{NOMENCLATURE} 

\begin{tabular}{ll}
$x$,$y$ &       coordinates on specimen surface in scan direction\\
        &       and perpendicular to this direction\\
$z$     &       coordinate along surface normal\\
$p_0$   &       equilibrium pressure\\
$R_0$   &       mean radius of surface corrugation\\
$T$     &       tensile strength of liquid\\
$Z_0$   &       amplitude of surface corrugation\\
\end{tabular}

\section{INTRODUCTION}

\begin{figure}[tbp]
\begin{center}
\includegraphics[width=0.8\columnwidth,bb=306 262 575 471]{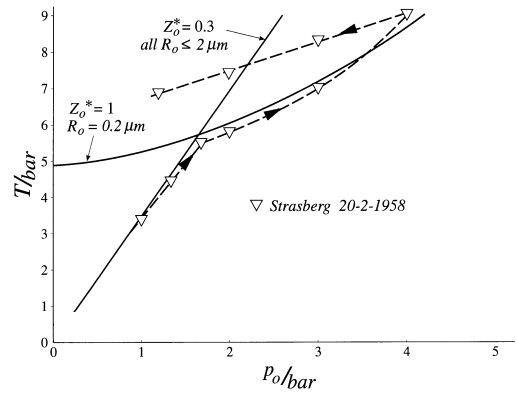}
\end{center}
\caption[]{Comparison between Strasberg's measurements ($\nabla$) of the
tensile strength $T$ of tap water vs. pressure $p_0$ \cite{ref5} and
calculations from \cite{ref4} based on large, shallow corrugations and
smaller, relatively deeper ones (solid lines).}
\label{FIG1}
\end{figure}

In cavitation research the formation and stabilization of
cavitation nuclei has always been an intriguing problem which
has made calculations of cavitation inception highly
problematic. Subcritical gas cavities in water are inherently
unstable and go into solution \cite{ref1} or they drift to surfaces due 
to buoyancy. Therefore, stabilization must take place at liquid-
solid interfaces. A model was proposed by Harvey \emph{et al.} \cite{ref2}.
Though able to explain some of the experimental results of
inception research, and during half a century the only
reasonably realistic model, it is insufficient. A new model was
proposed by M\o rch \cite{ref3}, and recently it has been improved to
allow quantitative calculations \cite{ref4}. According to this model
interfacial tension in the liquid adjacent to solid surfaces opens
the possibility of detachment of the liquid, i.e. void formation,
at surface elements of concave form. At sufficiently high
curvature the voids may develop spontaneously, but at
moderate and low curvatures the content of gas being in
solution in the liquid and reaching the interface by diffusion is
important for breaking liquid-solid bonds which are strained by
the interfacial tension in the liquid. It is predicted that a void
grows until the contact line between detached liquid and liquid
still in contact with the solid reaches the locus of balance
between the tensile stress due to interfacial liquid tension and 
the pressure in the bulk of liquid. It seems possible to explain
qualitatively the most significant results of experimental
research from this model. Quantitatively the measurements of
tensile strength $T$ of tap water vs. increasing equilibrium
pressure $p_0$ by Strasberg \cite{ref5} can be simulated from assuming
that solid particles in tap water have shallow corrugations of
sinusoidal cross section, axially symmetric around their bottom
and of mean radius $R_0 < 2\,{\rm \mu m}$ and with relative amplitude
$Z_0/R_0 = 0.3$, and small, relatively deeper ones with $R_0
=0.2\,{\rm \mu m}$,
$Z_0/R_0 =1$, FIG. \ref{FIG1}. 

\begin{figure}[tbp]
\begin{center}
\includegraphics[bb=179 511 424 618,width=0.88\columnwidth]{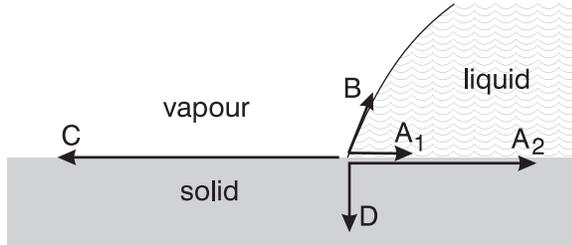}
\end{center}
\caption[]{Balance of interfacial forces at a solid-liquid-vapour
contact point. $A_1$ - interfacial liquid tension at liquid-solid
interface, $A_2$ - interfacial solid tension at liquid-solid interface,
$B$ - liquid-vapour interface tension, $C$ - solid-vapour interface
tension, $D$ - adhesive force.}
\label{FIG2}
\end{figure}

The interfacial tension present at the interface of two
substances in contact is normally given as a single quantity.
However, as the solid-liquid interfaces we consider are not
planar it is suitable to split this interface tension into two
components, one for the liquid, $A_1$, and one for the solid, $A_2$,
to obtain information of the influence of curvature on the
liquid-solid bonding. In FIG. \ref{FIG2} the balance of forces is shown
at a solid-liquid-vapour contact point. In addition to the
interface tension forces $A_1$, $A_2$, $B$ and $C$ at the three interfaces
this balance demands also an adhesion force $D$ (van der Waals'
force) between the liquid and the solid perpendicular to the
solid surface. These forces give the contact angle of the liquid-
vapour interface. In water where hydrogen bonds dominate the
intermolecular forces an appreciable interfacial liquid tension
($A_1$) is to be expected adjacent to solid surfaces as a result of
a stabilized interfacial liquid structure. Experimentally effects
of an orderly structured water layer have been measured near
a mica surface \cite{ref6} and by computer simulations it has been
shown that at platinum surfaces the interfacial layer of water
has an essentially ice-like solid structure \cite{ref7}. These results
support the hypothesis that water generally exhibits a more or
less stabilized structure at solid surfaces. The interfacial liquid
tension expected to result from this structure is the crucial
parameter in the model of void formation \cite{ref4} and thus for the
formation of cavitation nuclei in liquids. It is the object of the
present paper to verify its existence experimentally.     

\section{EXPERIMENTAL TECHNIQUE AND RESULTS} 

Experimental techniques available for investigating the local
interfacial tension in the liquid adjacent to a solid surface are
very few - at present only atomic force microscopy (AFM)
seems available \cite{ref8}. This technique is basically used to give
information of the surface topography of a solid object, and
resolution to atomic scale is available for crystallographically
planar surfaces. However, it can be used also for local force
spectroscopy. In AFM a pointed tip, usually of pyramidal form
and of height and base dimensions $5-10\,{\rm\mu m}$ and with a tip
radius of curvature $10-50\,{\rm nm}$, which is mounted close to the
free end of a thin cantilever of length about $300\,{\rm \mu m}$, is
approached to the surface which is to be investigated. When
the distance between the tip apex and the surface becomes
sufficiently small interatomic forces between the tip and the
object attract the tip, and the cantilever is bent. This is detected
by the deflection of a laser beam being reflected from the
cantilever surface opposite to the tip. The deflection is a
measure of the force on the tip. If the tip is approached further
to the surface contact is achieved and the resulting force shifts
into repulsion. This so-called contact mode is the one generally
used for topographic investigations. Here a suitable repulsive
deflection is chosen and the tip is scanned in the $x$- and $y$-
directions across the specimen while its height $z$ is regulated by
a feedback circuit to maintain the deflection chosen,
independent of surface corrugations. Thus the voltage in the
feedback circuit is a measure of the topographic changes.

In the force spectroscopy mode the tip is stationary in the $x$-
and $y$-directions, and the tip deflection is measured while the
cantilever base is moved along the $z$-axis at constant speed
towards the specimen until a suitable repulsive deflection is
achieved, then its motion is reversed. These investigations can
be made in vacuum, in gas, and in (optically transparent)
liquid. In vacuum only interatomic forces between tip and
specimen (van der Waals' forces) affect the deflection. In gas
(usually  atmospheric air) also forces between molecules
adsorbed to the surfaces are important. In particular water
molecules forming adsorbed water layers on the tip and
specimen surfaces are important because surface tension forces
cause strong attraction when these layers get in contact. The
surface tension forces and the van der Waals' forces result in
a transient "snap-in" of the tip at approch just before contact
is obtained. At operation in liquids it is generally assumed that
snap-in is absent because the liquid is taken to have bulk
structure right to the liquid-solid interface. The present results
indicate that this is not correct.

For the experiments a TopoMetrix TMX 2000 Explorer AFM
was used with V-shaped ${\rm Si}_3{\rm N}_4$ cantilevers of nominal spring
constant $0.03\,{\rm  N/m}$ and tip radius of curvature about $50\,{\rm nm}$. 
 
\begin{figure}[tbp]
\begin{center}
\includegraphics[bb= 171 420 416 521,width=0.88\columnwidth,clip]{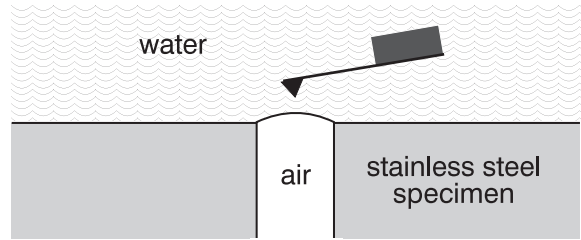}
\end{center}
\caption{Stainless steel specimen with central air-filled bore,
submerged in water. A submerged AFM tip and cantilever are positioned
above the water-air interface.}
\label{FIG3}
\end{figure}

The interfaces to be considered here are distilled water-air
interfaces which were approached from the liquid space, i.e.
with the tip and cantilever fully submerged, and diamond
polished stainless steel surfaces submerged in distilled water,
and an air-gold interface. The setup with cantilever and tip
submerged in water is shown in FIG. \ref{FIG3}. At suitable air
pressure in the central bore of the specimen a stable water-air
interface of form as a spherical segment is created at the top
of the bore, and it can be approached with the tip and
cantilever fully submerged in water. At lateral translation of
the specimen the water-stainless steel interface can be
investigated.  

\begin{figure}[tbp]
\begin{center}
\includegraphics[bb=341 93 544 356,width=0.9\columnwidth,clip]{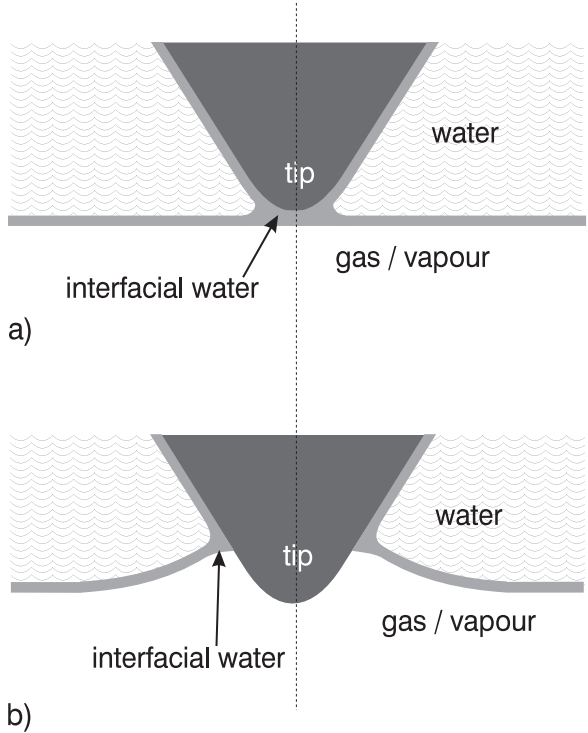}
\end{center}
\caption{a) Initial, non-equilibrium interaction between the orderly
structured interfacial layers of water at the AFM tip and a water-air
interface. b) Later stage, at which the water-air interface has moved
up along the tip surface while the tip itself is pulled down.}
\label{FIG4}
\end{figure}

When the tip approaches the water-air interface from the liquid
space and get in contact with the interfacial water it is strongly
attracted to the interface and crashes through it in a violent
snap-in. The process is interpreted to result from the interaction
of the orderly structured liquid at the water-air interface with
that at the water-tip interface. The initial interaction results in
an increased order in the zone of liquid around the tip apex,
and an attractive, but unbalanced force between the tip and the
water air-interface is set up by the interfacial liquid tension in
the structured zone, FIG. \ref{FIG4}a. A balance is then obtained by
local elevation of the water-air interface and bending of the
cantilever. As a consequence the tip breaks through the
interface, FIG. \ref{FIG4}b. With the soft cantilever used in the present
experiments balance was not achieved until the interface
reached the cantilever itself. It was not possible to record the
event as the dynamical range of the microscope ($\pm 7\,{\rm \mu m}$) was
greatly exceeded. 

\begin{figure}[tbp]
\begin{center}
\includegraphics[width=0.8\columnwidth,bb=314 499 573 828]{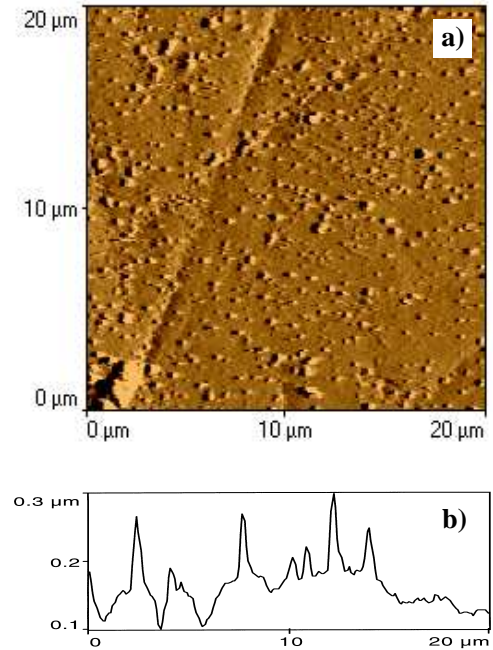}
\end{center}
\caption{a) Top view of the topography of a stainless steel surface as
recorded when submerged in water and observed in oblique
light. Top-bottom distance $500\,{\rm nm}$. b) Cross section along $x =
10\,{\rm \mu m}$.}
\label{FIG5}
\end{figure}

When subsequently the specimen was moved laterally to allow
investigation of the water-stainless steel interface the
topography of an area on the steel surface could be recorded
as shown in FIG. \ref{FIG5}a. A cross section along a single line, $x =
10\,{\rm \mu m}$, is shown in FIG. \ref{FIG5}b. The surface appears slightly
wavy with localized micro-hills in the $30-100\,{\rm nm}$ range. 

\begin{figure}[tbp]
\begin{center}
\includegraphics[bb=318 283 567 426,width=0.99\columnwidth,clip]{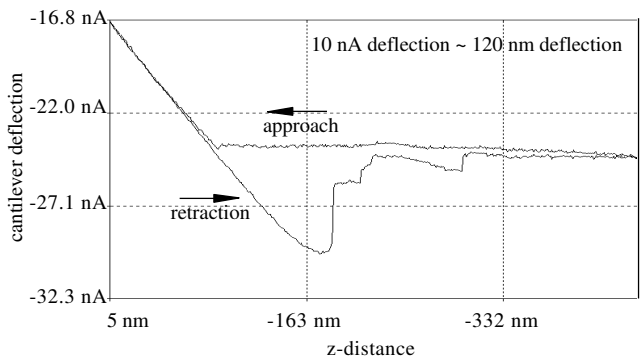}
\end{center}
\caption{Force spectroscopy on a water-stainless steel interface in a
case of small snap-in, but with notable effects of interfacial tension
at snap-out.}
\label{FIG6}
\end{figure}

By force spectroscopy it is found that the snap-in at approach
as well as the snap-out at the subsequent retraction depend
strongly on the location. Very often the sn
ap-in is quite small, just a few nm, and at retraction the tip
sticks to the solid surface until the cantilever base has retracted
about $100\,{\rm nm}$ corresponding to an attractive force of $3\,{\rm nN}$.
Then the tip escapes from the specimen surface, but it does not
return to the non-deflected condition until the cantilever base
has moved another $100\,{\rm nm}$ during which the tip relaxes in two
steps, FIG. \ref{FIG6}, found in repeated cases. This may be related to
the quantized adhesion reported in \cite{ref9}, though in the present
case the changes occur at a much larger scale. 

\begin{figure}[tbp]
\begin{center}
\includegraphics[bb=319 63 578 223]{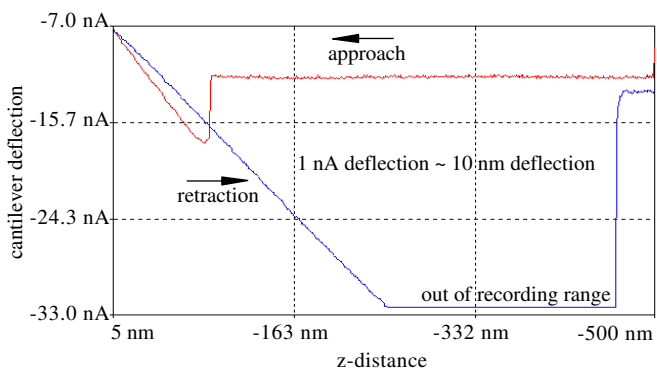}
\end{center}
\caption{Force spectroscopy on a water-stainless steel interface
showing a case of large snap-in (FIG.5a: $(x, y)\approx  (10\,{\rm \mu
m}, 10\,{\rm \mu m})$). A large retraction of the cantilever base
occurs before snap-out.} 
\label{FIG7}
\end{figure}

In other cases a very large snap-in occurs reproducibly, as
shown in FIG. \ref{FIG7}, where the snap-in is about $58\,{\rm  nm}$, and at
retraction the tip remains in contact until the cantilever base
has moved about $400\,{\rm nm}$. Then the cantilever returns to non-
deflected condition in a single jump.

The interpretation we give to these results is as follows: at
locations on the specimen surface where the liquid is in direct
contact with the solid surface there is an orderly structured
liquid layer of thickness about $1\,{\rm nm}$ adjacent to the solid, and
when the tip with its own orderly structured interface layer of
liquid, also of thickness about $1\,{\rm nm}$, approaches the solid
surface these interface layers merge and set up an attractive force on the tip, which in

\noindent combination with the van der Waals'
forces between tip and sample, being of range typically about
a few nanometer \cite{ref10}, result in a tip snap-in of less than
$10\,{\rm nm}$, as actually apparent from FIG. \ref{FIG6}. When the snap-in
brings the tip in contact with the solid surface the van der
Waals' forces are strongly enhanced. Therefore, a larger force
is required to withdraw the tip from contact. This is also
evident from FIG. \ref{FIG6} where retraction of the cantilever base
over a distance of about $100\,{\rm nm}$ is demanded to set up the
force needed for the tip to escape the surface itself. However,
it appears that a bending force on the cantilever remains. We
suppose this is a consequence of a nanovoid being formed
between the tip apex and the sample when contact between the
two solid surfaces is broken. The surface tension at the double
curved liquid-vapour interface which connects the tip and
sample and bounds the void prevents that liquid flows into the
gap and at the same time it establishes an attractive force
between tip and sample. Therefore, the tip is not totally free of
interaction with the specimen until the distance becomes so
large that the void collapses. This appears to happen after a
further $100\,{\rm nm}$ withdrawal, though in steps. The intermediate
jumps may be related to discontinuous changes of the loci of
contact of the liquid surface to the tip and sample surfaces.
Apparently this takes place when the force imposed by the bent
cantilever exceeds about $1-2\,{\rm nN}$. 

In cases of a significant snap-in, as in FIG. \ref{FIG7}, the event
cannot be attributed to neither contact between the structured
interfacial liquid at the tip and specimen surfaces nor to the
van der Waals' forces, as these do not extend beyond at most
$10\,{\rm nm}$. In liquid the range of the van der Waals' forces is
actually reduced compared to their range in air \cite{ref11}. However,
if a stable interfacial void has grown on the specimen surface
due to the local characteristic features of this surface, as
modelled in \cite{ref4}, the tip meets a water-gas interface during the
approach. As described above such an interface attracts the tip
strongly and makes it penetrate deeply, i.e. in the present case
it penetrates until tip-solid contact prevents further penetration,
and a significant repulsive force between tip and specimen
surface may then occur. This interpretation is supported by the
large retraction distance of $400\,{\rm nm}$, corresponding to an
attractive force of $12\,{\rm nN}$, observed before snap-out occurs, and
the tip now escapes the solid surface as well as the supposed
surface-attached void in a single large jump. This force
considerably exceeds the van der Waals' forces on the tip,
which must be smaller than the $3\,{\rm nN}$ found from FIG. \ref{FIG6}.
Thus it reveals strong interfacial forces in the liquid adjacent
to a water steel interface.

\begin{figure}[tbp]
\begin{center}
\includegraphics[bb=123 575 384 705]{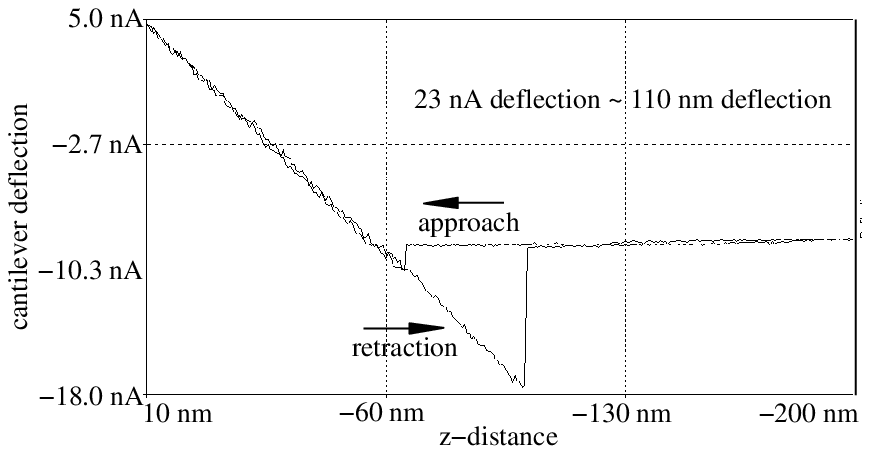}
\end{center}
\caption{Force spectroscopy on an air-gold interface.}
\label{FIG8}
\end{figure}

It is of interest to compare the above results from stainless
steel surfaces submerged in water with observations in air of
a solid surface which does not adsorb water. In such a case the
water adsorbed to the tip is of no significance as water is not
attracted to the non-adsorbing solid surface. Gold does not
adsorb water to any significant extent, and in FIG. \ref{FIG8} force
spectroscopy on an air-gold interface is shown. The snap-in is
about $9\,{\rm nm}$, and it can be ascribed to van der Waals' forces
which are unscreened due to the absence of water on the gold
surface. At retraction the snap-out occurs in a single jump after
$43\,{\rm nm}$ withdrawal of the cantilever base, corresponding to van
der Waal's forces of only $1.3\,{\rm nN}$. 

If we compare the force spectroscopy of the submerged stain-
less steel surface in contact with water, FIG. \ref{FIG6}, with that of
the air-gold interface, we notice that at the submerged water-
stainless steel interface the van der Waals' forces are notably
weaker at snap in, probably due to the screening effect of
water. At snap out however, the upper limit of the van der
Waals' forces can be estimated from the air-gold experiment,
and they are clearly insufficient to explain the force needed for
the first snap-out. Therefore, by the first snap-out in FIG. \ref{FIG6}
already a major part of the attractive force can be ascribed to
liquid interfacial tension.

\section{CONCLUSION}

AFM force spectroscopy investigations at solid as well as
gaseous interfaces with water, probed from the liquid space,
reveal characteristic attractive forces which can only be
attributed to liquid tension in the interfacial water. This brings
experimental support to the model of void formation at liquid-
solid interfaces \cite{ref4} in which the interfacial liquid tension is a
basic assumption. Further, the presence of interfacial voids is
actually experimentally supported. Such voids are sources of
cavity formation when single-phase liquids are exposed to
tensile stress.

\newpage

\widetext

\end{document}